\documentclass[onecolumn,floatfix,superscriptaddress,showpacs,showkeys,nofootinbib,preprint]{revtex4}
\usepackage{epsfig}
\usepackage{amssymb,latexsym,amsmath}
\newcommand{\eq}[1]{\begin{align} #1 \end{align}}

\begin{document}
\title{Modified Bag Models for   \\ the Quark Gluon Plasma Equation of State
}

 \author{V. V. Begun}
 \affiliation{Bogolyubov Institute for Theoretical Physics,
 Kiev, Ukraine}
 \affiliation{Frankfurt Institute for Advanced Studies, Frankfurt,
 Germany}

 \author{M. I. Gorenstein}
 \affiliation{Bogolyubov Institute for Theoretical Physics,
 Kiev, Ukraine}
 \affiliation{Frankfurt Institute for Advanced Studies, Frankfurt, Germany}

 \author{O. A. Mogilevsky}
 \affiliation{Bogolyubov Institute for Theoretical Physics,
 Kiev, Ukraine}


\begin{abstract}
The modified versions of the bag model equation of state (EoS) are
considered. They are constructed to satisfy the main qualitative
features observed for the quark-gluon plasma EoS in the lattice
QCD calculations. A quantitative comparison with  the lattice
results at high temperatures $T$ are done in the SU(3) gluodynamics and in the full QCD with
dynamical quarks.
Our analysis advocates
a negative value of the bag constant $B$.
 \end{abstract}



\maketitle

\section{Introduction}

A transition to the deconfined phase of quarks and gluons, the
quark gluon plasma (QGP), is expected at high temperature and/or
baryonic density (see, e.g., Refs.~\cite{Kapusta} and \cite{Satz}
and references therein).
In the present study of the QGP equation of state (EoS) we
consider the system with zero values of all conserved charges.
This is approximately valid for the QGP created in nucleus-nucleus
collisions at the BNL RHIC and even better for future experiments
at the CERN LHC.
Up to now the  strongly interacting matter EoS could be only
calculated from the first principles within the lattice QCD. These
calculations are done for zero or very small values of the
baryonic chemical potential. The QGP  exists at high temperatures
$T>T_c$, where the critical temperature $T_c$ corresponds to the
$1^{\text{st}}$ order  phase transition in the pure SU(3)
gluodynamics or to a smooth crossover in the full QCD. The main
results for the QCD deconfined matter EoS can be illustrated by
the Monte Carlo (MC) lattice results~(LR)  for the energy density
$\varepsilon(T)$ and pressure $p(T)$ in the SU(3) gluodynamics
\cite{lattice}.
%
%
The qualitative features of the EoS at $T>T_c$ can be summarized
as follows. The pressure $p(T)$ is very small at the critical
temperature, $p(T_c)/T_c^4<< 1$, and rapidly increases at
$T\gtrsim T_c$~. At high $T$ the system reaches the ideal massless
gas behavior $p \cong \varepsilon/3$, thus, $\varepsilon(T)\cong
\sigma T^4$~. However, the constant $\sigma$ which regulates the
high temperature behavior is about $10\%$ smaller than the
Stefan-Boltzmann (SB) constant $\sigma_{SB}$~. Both
$\varepsilon/T^4$ and $3p/T^4$ approach the value $\sigma $ from
below. The interaction measure $ (\varepsilon -3p)/T^4$, called
also the trace anomaly, demonstrates a prominent maximum at
$T\cong 1.1~T_c$~. Note that  these properties of the gluon plasma EoS
are also valid in the full QCD  \cite{LQCD,LQCD1,q-lattice} .

%

%
The bag model (BM) \cite{BM} was invented to describe the mass
spectrum of the hadron states. Soon after that it was suggested
\cite{baacke} to interpret the bag constant $B$ as the
non-perturbative energy density term in the deconfined matter EoS.
For several decades, the BM EoS has been used to describe the QGP
(see, e.g., Ref.~\cite{bag}). In its simplest form, i.e. for
non-interacting massless constituents and zero values of all
conserved  charges, the BM EoS reads:
\eq{\label{bag-a}
\varepsilon(T)~=~\sigma_{SB}~T^4~+~B~,
~~~~~~p(T)~=~\frac{\sigma_{SB}}{3}~T^4~-~B~,
}
where $\varepsilon$ and  $p$  have a simple dependence on $T$
modified by adding the bag constant $B$~(``vacuum pressure'').
The SB constant in Eq.~(\ref{BM}) is
$\sigma_{SB}= \pi^2/30~ \left( d_B+7d_F/8\right) $, where $d_B$
and $d_F$ are the degeneracy factors for the massless bosons
(gluons) and fermions (quarks and anti-quarks), respectively.

The main goal of the present paper is to study the modifications
of the bag model EoS. We consider simple analytical
parameterizations for the QGP EoS which include a linear and/or
quadratic in $T$ terms in the pressure function to satisfy the
qualitative properties listed above. The quantitative comparison
with the MC LR in the SU(3) gluodynamics \cite{lattice} and in the
full QCD with dynamical quarks \cite{q-lattice} will be done in
Sections II and III, respectively. The Section~IV summarizes the
paper.

\section{Gluon plasma Equation of State}
In this Section a quantitative comparison of the modified versions
of the BM EoS is done with the MC LR \cite{lattice} in the pure
SU(3) gluodynamics.  The $\varepsilon/T^4$ and $3p/T^4$ were
obtained in Ref.~\cite{lattice} by extrapolation to an infinite
continuous system. We take the MC values of these extrapolated
functions $\varepsilon/T^4$  and $3p/T^4$ at the same $T/T_c$
points where the interaction measure $(\varepsilon -3p)/T^4$ has
been simulated on the finite lattice  $32^3\times 8$~. To
determine the parameters of different models discussed below we
will minimize the sums of the square deviations at these $T/T_c$
points for $(\varepsilon - \varepsilon_{MC})/T^4$ and/or
$3(p-p_{MC})/T^4$, where $\varepsilon$,  $p$ are the model
functions and $\varepsilon_{MC}$, $p_{MC}$ are the MC LR.


\vspace{0.2cm} The recent lattice estimate for the pressure at
very high temperatures $T/T_c\cong 10^7$ is still about $3\%$
below the SB limit \cite{Fodor}. The lowest order perturbative
calculations give $(\sigma_{SB}-\sigma) \propto g^2(T)\propto
1/\ln(T/\Lambda)$.
The calculations within the perturbative re-summation scheme
\cite{blaizot} are comparable with the LR at $T= (3\div4)~T_c$~
and suggest that the dominant effect of interactions is to turn
massless quarks and gluons into weakly interacting quasiparticles.
The quasiparticle approach of Ref.~\cite{GY} (see also recent
papers \cite{rev} and references therein) treats the system of
interacting gluons as a gas of non-interacting quasiparticles with
gluon quantum numbers, but with mass $m(T)$ which depends on $T$.
The particle energy $\omega$ and momentum $k$ are assumed to be
connected as
%
%
$\omega = \left[k^{2}~+~m^{2}(T)\right]^{1/2}$~.
%
The energy density and pressure take then the following form
\cite{GY}:
 \eq{
 \varepsilon(T)~&=~\frac{d}{2\pi^{2}}\int_{0}^{\infty}k^2 dk
~\frac{\omega}{\exp(\omega/T)~-~1}~
+~B^*(T)~\equiv~\varepsilon_0(T,\omega)~+~B^*(T)~,\label{epsilon}
\\
 p(T)~&=~\frac{d}{6\pi^{2}}~ \int_{0}^{\infty} k^2 dk
~\frac{k^{2}}{\omega}~\frac{1}{\exp(\omega/T)~-~1}
~-~B^*(T)~\equiv~p_0(T,\omega)~-~B^*(T)~,\label{pressure}
}
where the degeneracy factor $d=2(N_{c}^{2}-1)$ equals 16 for the
SU(3) gluodynamics. The thermodynamical relation,
\eq{\label{therm}
T\frac{dp}{dT}~-~p(T)~=~\varepsilon(T)~,
}
leads to the equation for the function $B^*(T)$,
\eq{\label{BT}
 ~\frac{dB^*}{dT} ~=~ -~\frac{\Delta_0(T,\omega)}{m}~\frac{dm}{dT}~,
}
where $ \Delta_0\equiv \varepsilon_0~-~3p_0$, and $\varepsilon_0$,
$p_0$ defined by Eqs.~(\ref{epsilon},\ref{pressure}) are the ideal
gas expressions for massive bosons. If the function $m(T)$ is
known one can calculate $B^*(T)$ from Eq.~(\ref{BT}) up to an
arbitrary integration constant $B$.
%
For
$m=aT$, where $a$ is a temperature independent parameter, the
function $B^*(T)$ derived from Eq.~(\ref{BT}) equals to
\cite{BGM}:
\eq{\label{BT2}
B^*(T)~=~B-~\frac{1}{4}~\Delta_0(T,\omega)~.
}
One obtains the energy  density (\ref{epsilon}) and the pressure
(\ref{pressure}),
\eq{\label{BM}
 \varepsilon(T)~=~\sigma~T^4~+~B~,~~~~~~ p(T)~=~\frac{\sigma}{3}~T^4~-~B~,
}
which has the form of the BM (\ref{bag-a}) with  constant $\sigma$
equal to
 \eq{ \label{sigma1}
 \sigma~ = ~{3d\over 2\pi^{2}}
~\sum_{n=1}^{\infty}\left[ {a^{2}\over n^{2}}~K_2(na)~ +~{a^3\over
4n}~
 K_{1}(na)\right] ~\equiv~ \kappa(a)~\sigma_{SB}~.
}
The $K_{1}$ and $K_{2}$ in Eq.~(\ref{sigma1}) are the modified
Bessel functions. The constant $\sigma$ in Eq.~(\ref{BM}) includes
the suppression factor $\kappa(a)$. For $a\rightarrow 0$, it
follows $\kappa \rightarrow 1$, and Eq.~(\ref{BM}) coincides with
Eq.~(\ref{bag-a}).  The function $\kappa(a)$ decreases
monotonously and goes to zero at $a\rightarrow \infty$.


%

\begin{figure}[ht!]
\epsfig{file=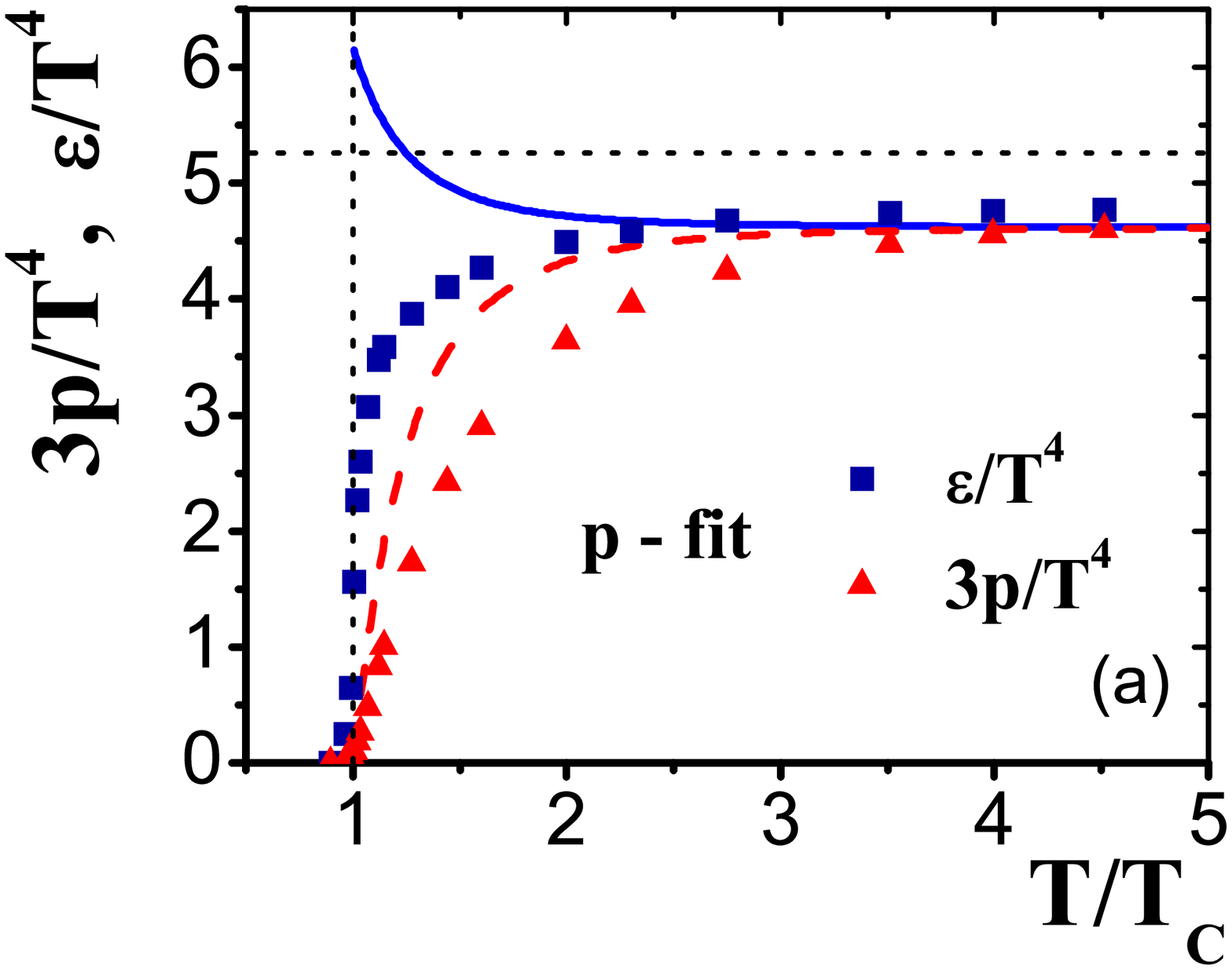,width=0.49\textwidth}~~
\epsfig{file=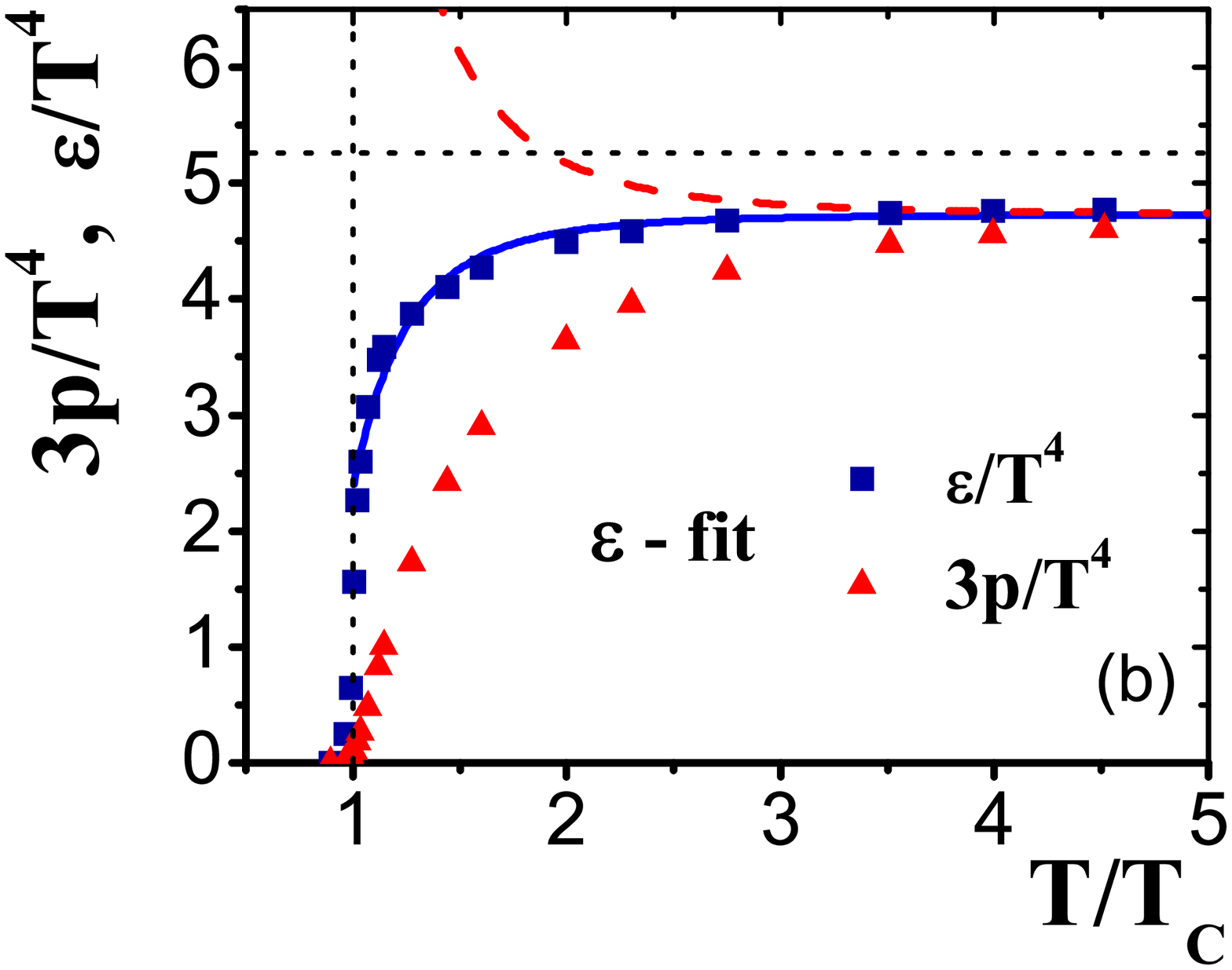,width=0.49\textwidth}
\caption{(Color online) The  MC LR for the SU(3) gluodynamics. The
$\varepsilon/T^4$ (squares) and $3p/T^4$ (triangles) are
extrapolated to infinite continuous system \cite{lattice}. The
dotted vertical and horizontal lines correspond to $T/T_c=1$ and to the Stefan-Boltzmann
constant $\sigma_{SB}=8\pi^2/15$, respectively.
%
%
%
The dashed lines show $3p/T^4$ and solid ones $\varepsilon/T^4$
for the BM EoS (\ref{BM}).~~ {\bf a}:~The fit of $3p/T^4$ with the
BM EoS (\ref{BM}) gives $\sigma=4.62$ and  $B= 1.56~ T_c^4$.~~
{\bf b}:~The fit of $\varepsilon/T^4$ with the BM EoS (\ref{BM})
gives $\sigma=4.73$ and $B= -~2.37~ T_c^4$~. \label{fig-bag}}
\end{figure}

\vspace{0.2cm} The modified SB constant $\sigma < \sigma_{SB}$
allows to fit the high temperature behavior of $\varepsilon(T)$
and $p(T)$. In what follows the BM EoS (\ref{BM}) is considered
with $B$ and $\sigma$ being free model parameters.  The LR
\cite{lattice} cover the temperature range $(0.89\div4.5)T_c$. We
consider the high temperature phase (``gluon plasma'') at $T>T_c$,
where $T_c$ is a point of the 1$^{\text{st}}$  order phase
transition. To be precise, let us note that we use the MC LR
\cite{lattice} for $T>1.02~T_c$ to avoid the uncertainties at $T=
T_c$  where $\varepsilon(T)$ has a discontinuity in
thermodynamical limit. The fit of the MC LR for $3p/T^4$
gives $\sigma=4.62$ and $B=1.56~T_c^4$, and it is shown by the
dashed line in Fig.~\ref{fig-bag}a. One observes a correct
behavior,  $3p/T^4\cong \sigma$, at high $T$ and an abrupt drop
near the critical temperature, $3p(T_c)/T_c^4 \approx 0$~. These
features of $p(T)$ are in a qualitative agreement with the LR.  A
quantitative agreement is however unsatisfactory. Moreover, the
temperature dependence of $\varepsilon/T^4$ calculated by
Eq.~(\ref{BM}) with $\sigma=4.62$ and $B=1.56~T_c^4$ appears to be
in a qualitative contradiction with the MC LR (see the solid line
in Fig.~\ref{fig-bag}a).

One can alternatively start from fitting the MC LR for the energy
density function $\varepsilon(T)$ with Eq.~(\ref{BM}).
Unexpectedly, one obtains a rather good agreement with MC LR for
$\varepsilon/T^4$ admitting {\it negative} values of the bag
constant $B$~. The negative bag constant $B=-~2.37~T_c^4$ and
$\sigma=4.73$ needed in Eq.~(\ref{BM}) to fit $\varepsilon/T^4$
leads, however, to an incorrect behavior of $p/T^4$ (see the
dashed line in Fig.~\ref{fig-bag}b).

\begin{figure}[ht!]
\epsfig{file=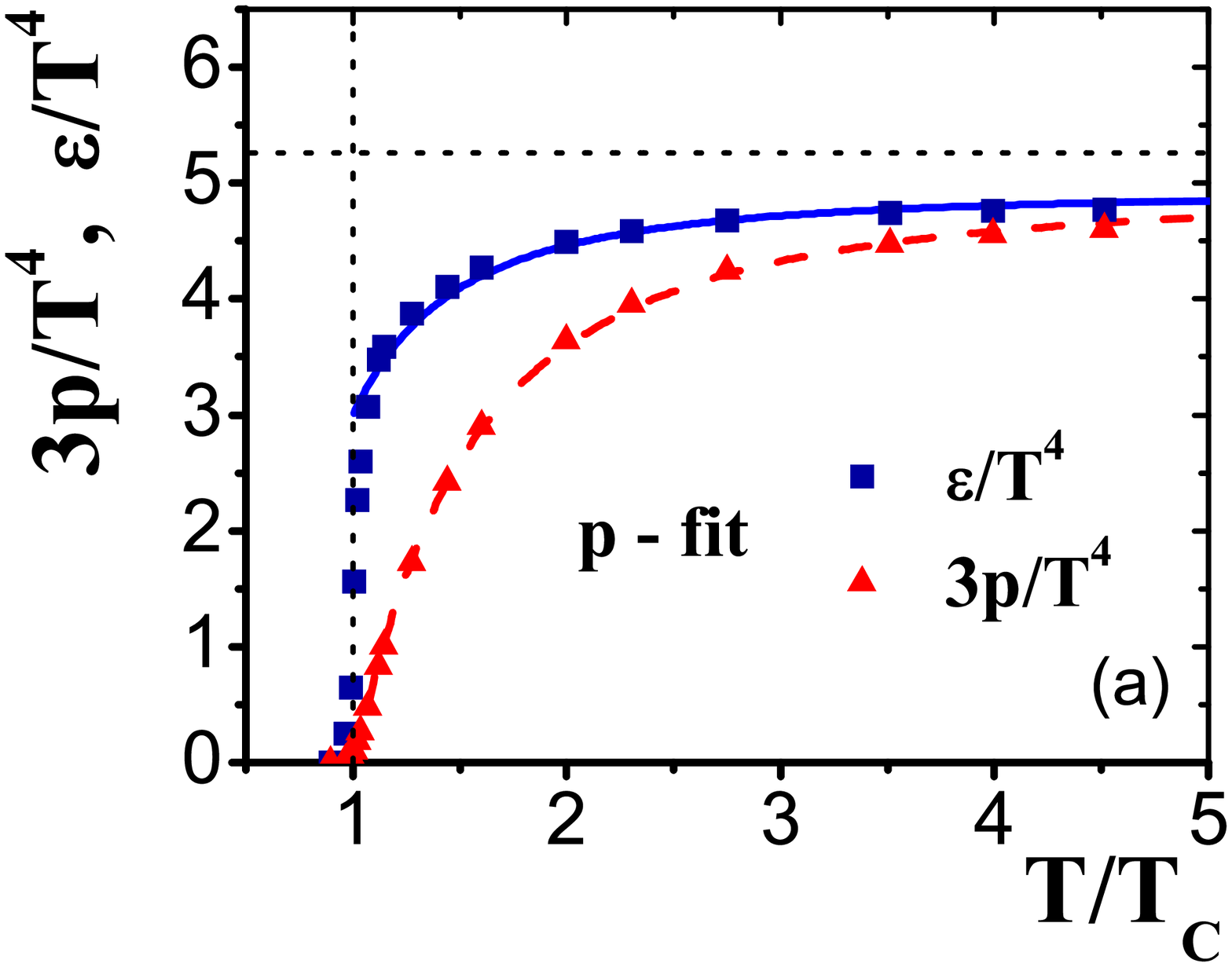,width=0.49\textwidth}~~
\epsfig{file=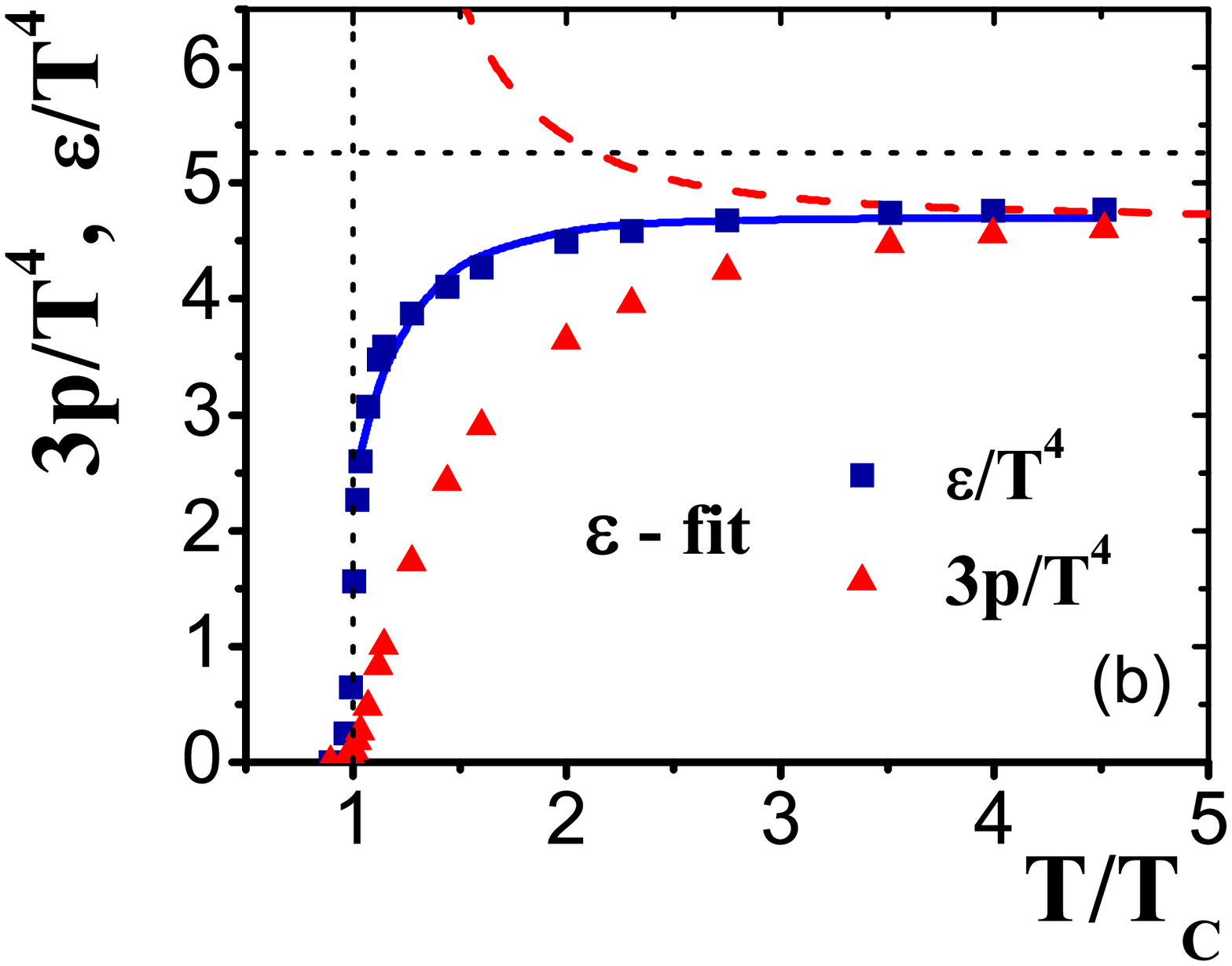,width=0.49\textwidth}
\caption{(Color online)
 The MC LR \cite{lattice} are the same as in Fig.~1.
 The dashed lines show $3p/T^4$ and solid ones
$\varepsilon/T^4$ for the C-BM EoS (\ref{CBM}).~~ {\bf a}:~The fit
of $3p/T^4$ with the C-BM EoS (\ref{CBM}) gives $\sigma=4.92$,
$B=-~0.13~T_c^4$, and $C=1.8~T_c^2$.~~ {\bf b}:~The fit of
$\varepsilon/T^4$ with the C-BM EoS (\ref{CBM}) gives
$\sigma=4.69$, $B=-~2.64~T_c^4$, and $C=-~0.28~T_c^2$~.
 \label{fig-BC} }
\end{figure}

\vspace{0.2cm}
%
A modification of the BM EoS (\ref{BM}) was considered by
Pisarski \cite{Pisarski}:
\eq{\label{CBM}
\varepsilon(T)~=~\sigma~T^4~-~C~T^2~+~B~, ~~~~p(T)~=~
\frac{\sigma}{3}~T^4~-~C~T^2~-~B~.
}
A presence of the  $T^2$-terms in $p(T)$ and $\varepsilon(T)$ has
been further studied in recent papers \cite{T2}. For brevity, we
will refer to  Eq.~(\ref{CBM}) as the ``C-bag model'' (C-BM). The
fit of the MC LR for  $p/T^4$  with the C-BM EoS (\ref{CBM}) is
presented in Fig.~\ref{fig-BC}a.  It gives, $\sigma=4.92$,
$B=-~0.13~T_c^4$, and $C=1.8~T_c^2$~. One finds an agreement of
the C-BM EoS (\ref{CBM}) with the LR for $3p/T^4$~. In particular,
$3p(T_c)/T_c^4 \approx 0$~. However,
$\varepsilon(T_c)/T_c^4\approx 3$, which exceeds the LR.

Trying to improve the quantitative agreement with the LR, one may
start from fitting the $\varepsilon/T^4$  with Eq.~(\ref{CBM}).
One observes indeed a better agreement for $\varepsilon/T^4$ with
the parameters  $B=-~2.64~T_c^4$ and $C=-~0.28~T_c^2$, being very
different from those found in the fitting of $3p/T^4$~. These new
values of $B$ and $C$ lead, however, to a qualitative disagreement
of Eq.~(\ref{CBM}) with $3p/T^4$ LR, as shown by the dashed line
in Fig.~\ref{fig-BC}b.

\begin{figure}[ht!]
\epsfig{file=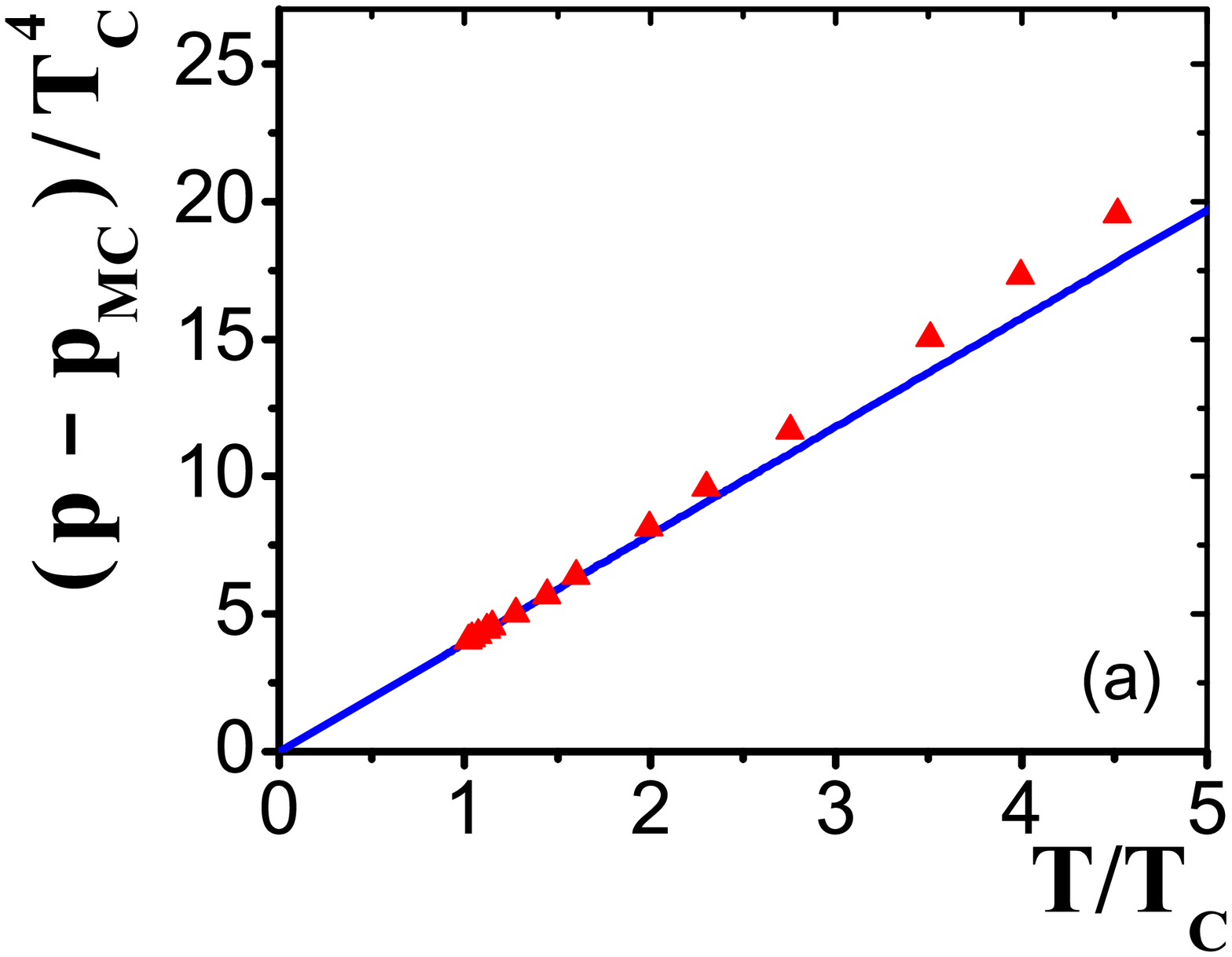,width=0.49\textwidth}~~
\epsfig{file=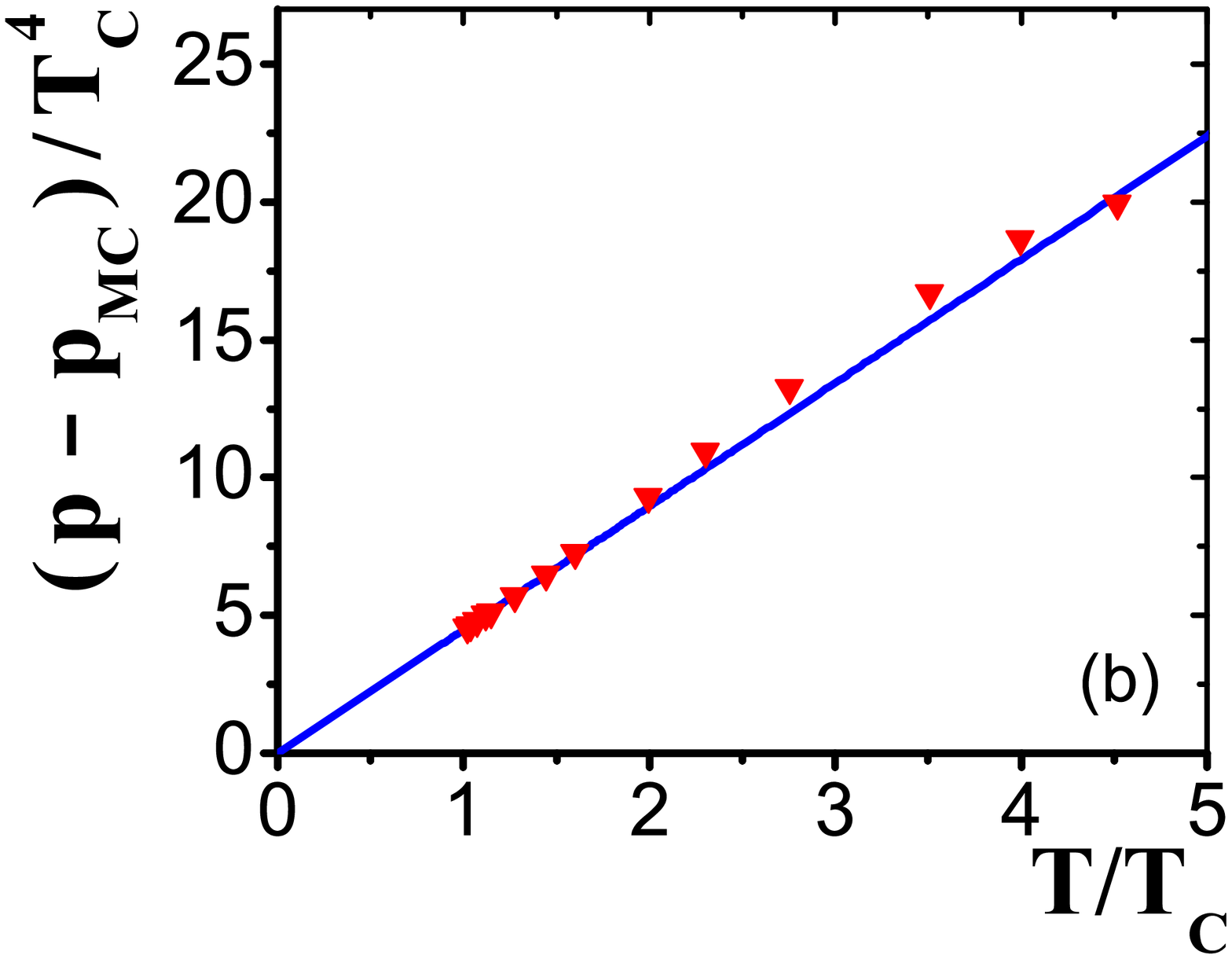,width=0.49\textwidth}
\caption{(Color online) A difference of the model pressure $p$ and
the MC LR $p_{MC}$ divided by $T_c^4$.~~ {\bf a}:~Pressure $p$ is
given by the BM EoS (\ref{BM}) with  $\sigma=4.73$ and
$B=-~2.37~T_c^4$~. The Solid line presents the linear function
$3.94~ T/T_c$.~~ {\bf b}:~Pressure $p$ is given by the C-BM EoS
(\ref{CBM}) with $\sigma=4.69$, $B=-~2.64~T_c^4$, and
$C=-~0.28~T_c^2$~. The solid line presents the linear function
$4.48~ T/T_c$~. \label{fig-AB-1} }
\end{figure}

\vspace{0.2cm}
%
Comparing the BM EoS (\ref{BM}) and C-BM EoS (\ref{CBM}) with the
MC LR we have faced the serious challenge. Very different values
of model parameters, $B$ for Eq.~(\ref{BM}), or $B$ and $C$ for
Eq.~(\ref{CBM}), have been found depending on whether we start
from fitting $3p/T^4$ or from $\varepsilon/T^4$~. By admitting
negative values of the bag constant $B$ in Eq.~(\ref{BM}) or
Eq.~(\ref{CBM}), one obtains a good fit of $\varepsilon/T^4$ in
the whole temperature interval $T>T_c$, but finds a  disagreement
with LR for $3p/T^4$,  as seen from Fig.~\ref{fig-bag}b and
Fig.~\ref{fig-BC}b. This finding looks contra-intuitive in view
that the functions $\varepsilon(T)$ and $p(T)$ are in the
one-to-one correspondence to each other due to the thermodynamical
consistency equation (\ref{therm}).
In Fig.~\ref{fig-AB-1} we show the differences between the
pressure functions $p(T)$ calculated in the BM EoS (\ref{BM})
or in the C-BM (\ref{CBM}),with parameters obtained from the fit
of $\varepsilon/T^4$, and the MC LR for pressure $p_{MC}(T)$~. The
difference of the pressures is divided by $T_c^4$~.
From
Fig.~\ref{fig-AB-1} one clearly observes a linear temperature
dependence of $(p-p_{MC})/T_c^4$~.
%
The thermodynamical relation (\ref{therm}) does connect the
functions $\varepsilon(T)$ and $p(T)$~. This connection is,
however, not symmetric in the two directions. If the function
$p(T)$ is known, one finds $\varepsilon(T)$ from Eq.~(\ref{therm})
in a unique way. However, if the function $\varepsilon(T)$ is
known, Eq.~(\ref{therm}) is the $1^{\text{st}}$  order
differential equation for the function $p(T)$~. The general
solution of this equation involves an arbitrary integration
constant. This results in a linear in temperature term in the
function $p(T)$~. Thus, for $\varepsilon(T)$ in the form of
Eq.~(\ref{BM}), a general solution of Eq.~(\ref{therm}) for $p(T)$
can be written as follows:
\eq{\label{ABM}
 \varepsilon(T)~=~\sigma~T^4~+~B~,~~~~~~
p(T)~=~\frac{\sigma}{3}~T^4~-~B~-A~T~.
}
The term $-A\,T$ with an arbitrary constant $A$ corresponds to a
general solution of the homogeneous equation $Tdp/dT - p=0$~ as
was noticed in Refs.~\cite{Kallman,GM}.
For brevity we call the EoS
(\ref{ABM}) the ``A-bag model'' (A-BM).
%
%
The A-BM EoS (\ref{ABM}), in contrast to the BM EoS (\ref{BM}) and
C-BM EoS (\ref{CBM}), gives essentially the same values of the
model parameters $\sigma,~ B,$ and $ A$ either one starts from
fitting $3p/T^4$ or from $\varepsilon/T^4$~.
\begin{figure}[ht!]
\epsfig{file=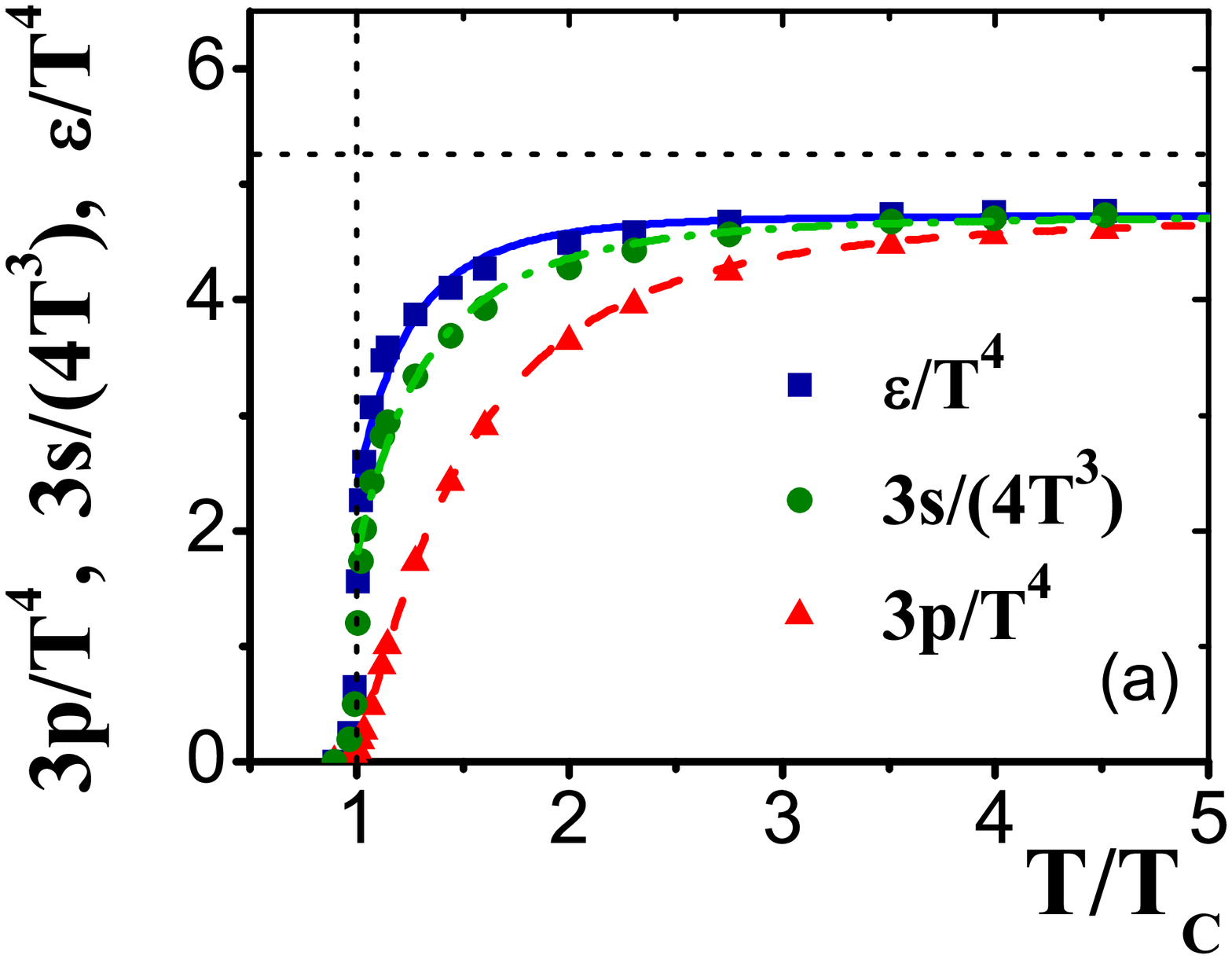,width=0.49\textwidth}
\epsfig{file=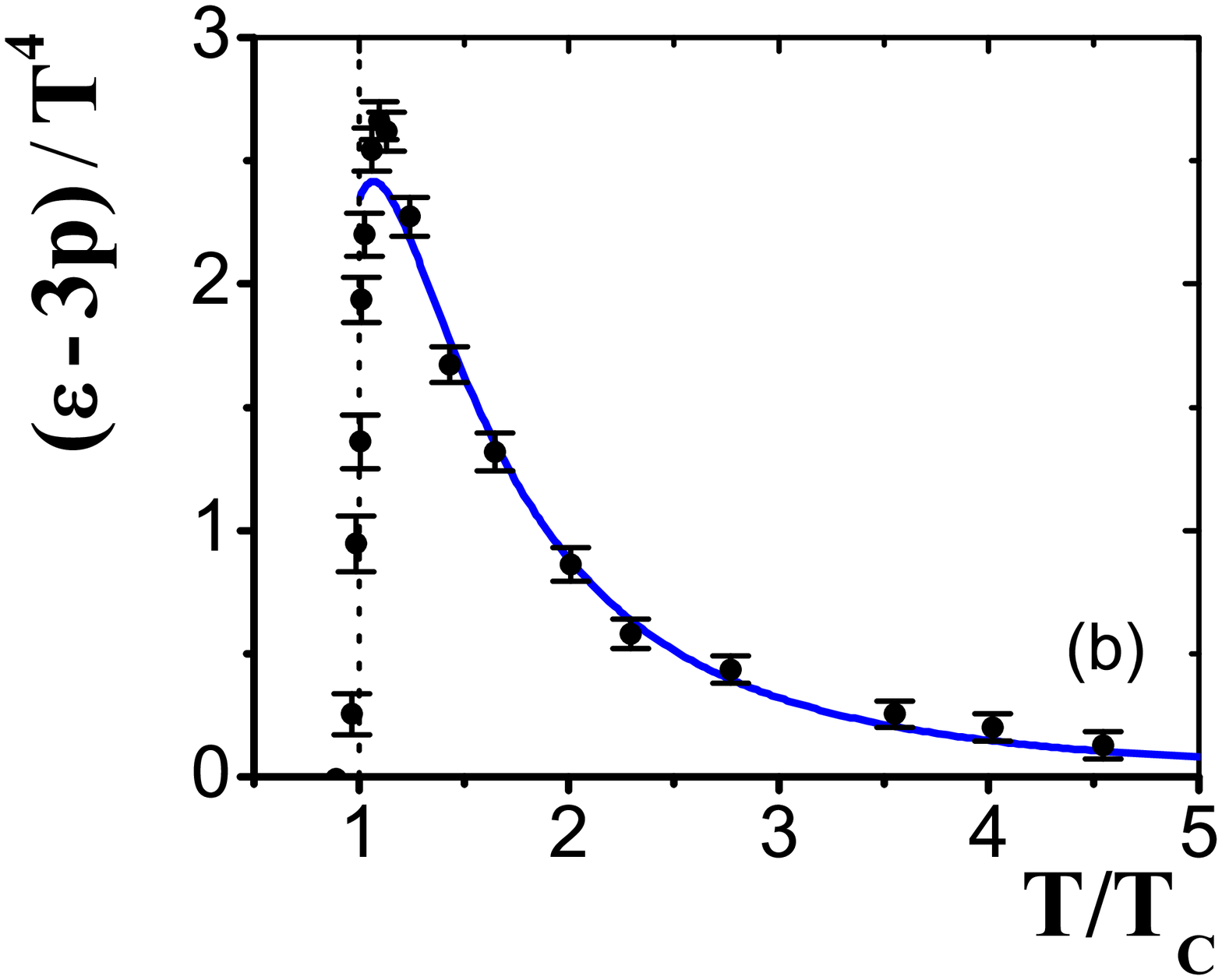,width=0.49\textwidth}
\epsfig{file=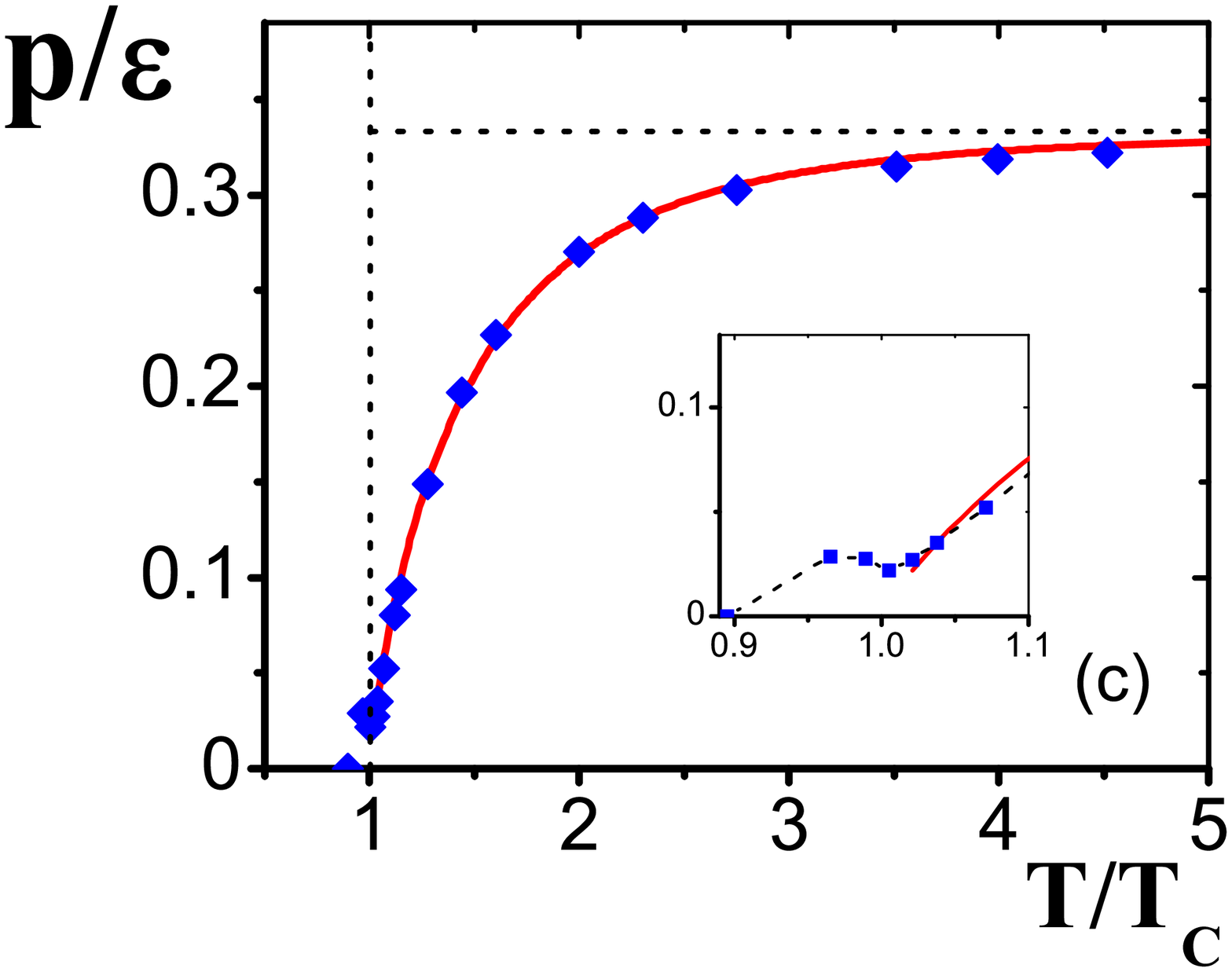,width=0.49\textwidth}
 \epsfig{file=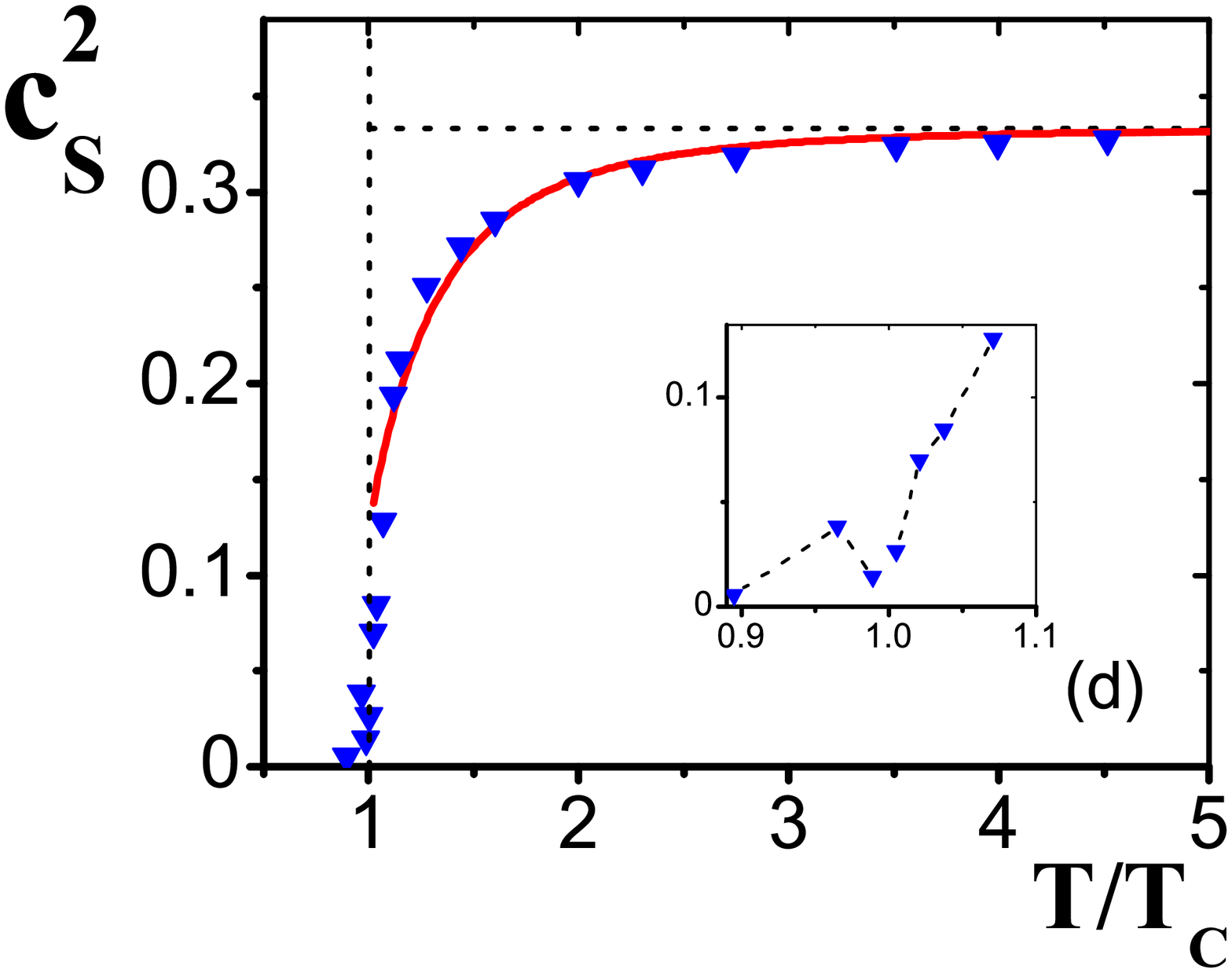,width=0.49\textwidth}
 \caption{(Color online) The symbols show the
MC LR in the SU(3) gluodynamics \cite{lattice}. The lines
correspond to the A-BM EoS (\ref{ABM}) with $\sigma
=4.73,~A=3.94~T_c^3,~B=-~2.37~T_c^4$. The dotted horizontal lines
in figures {\bf c} and {\bf d} correspond to $p/\varepsilon =1/3$ and
$c_s^2=1/3$, respectively. ~~ {\bf a}:~The squares are
$\varepsilon/T^4$, triangles $3p/T^4$, and circles $3s/(4T^3)$.~~
{\bf b}:~The interaction measure $ (\varepsilon -3p)/T^4$.~~
{\bf c}:~The ratio $p/\varepsilon$.~~ {\bf d}:~The speed of sound
squared $c_s^2$.
 \label{fig-delta1} }
\end{figure}
Figure \ref{fig-delta1} demonstrates a good agreement of the A-BM
(\ref{ABM}) with the MC LR for the thermodynamical functions
$\varepsilon/T^4$, $3p/T^4$, and $3s/(4T^3)$ (where
$s=(\varepsilon + p)/T$ is the entropy density), interaction
measure, $(\varepsilon -3 p)/T^4$, the ratio $p/\varepsilon$, and
speed of sound squared, $c_s^2=dp/d\varepsilon$. For the A-BM EoS
(\ref{ABM}), $(\varepsilon -3 p)/T^4$ does not depend on the
parameter $\sigma$ whereas the entropy density $s(T)$ does not
depend on the bag parameter $B$.

Let us consider the EoS which includes both $C\,T^2$ and $A\,T$
terms,
\eq{\label{ABC}
\varepsilon(T)~=~\sigma~T^4~-~C~T^2~+~B~,~~~~~
p(T)~=~\frac{\sigma}{3}T^4~-~C~T^2~-~A~T~-~B~,
}
referred to as the AC-BM. The standard BM EoS (\ref{BM})
corresponds to $A=C=0$, whereas the C-BM EoS (\ref{CBM}) and A-BM
EoS (\ref{ABM}) correspond to $A=0$ and $C=0$ in Eq.~(\ref{ABC}),
respectively.
%
%
%
%
%
%
%
%
A comparison of the AC-BM EoS (\ref{ABC}) with the MC LR for the
$\varepsilon/T^4$ and $3p/T^4$ in SU(3) gluon plasma leads to
$C/T_c^2<<1$~.
We thus conclude that the AC-BM ({\ref{ABC}) for the gluon plasma
is reduced to the A-BM EoS (\ref{ABM}).
Note that the lattice study of SU(N$_c$) gluodynamics with
$N_c=3,~4,~5,~6,$ and 8 colors performed in Ref.~\cite{Nc} reveals that
$3p/T^4$ and $\varepsilon/T^4$ divided by the corresponding SB limits
follow essentially the same curves for different N$_c$~.

%
A physical origin of the linear in $T$ term requires further
studies. In this connection we remind the famous problem of Gribov
copies \cite{gribov} and his suggestion of the modified gluon
dispersion relation,
%
%
$\omega(k)~=~\sqrt{k^2~+~M^4/k^2}$~,
%
%
where $M$ is a QCD mass scale.
It was shown in Ref.~\cite{zwan} that at $T>>M$ this dispersion
relation gives the SB limit $\varepsilon/T^4=3p/T^4=\sigma_{SB}$
and power corrections of relative order $1/T^3$ for $p/T^4$ and
$1/T^4$ for $\varepsilon/T^4$~.
This also resembles the cut-off $K$ phenomenological model
\cite{K} where 
$\omega(k)=k~\theta(k-K)$~, i.e.
low-momentum gluons are suppressed but high-momentum gluons are
effectively free.

\section{Quark Gluon Plasma Equation of State}

The LR for the realistic equation of state -- the QCD with 2+1
flavors (light $u$-, $d$- and  heavier $s$-quarks) have been
presented by ``HotQCD'' \cite{LQCD} and ``Wuppertal-Budapest''
\cite{LQCD1,q-lattice} collaborations.
We compare the modified BM EoS with the latest LR
\cite{q-lattice}. The continuum estimates of the LR results for
$p/T^4$, $(\varepsilon -3p)$, and $c_s^2$ in the temperature range
100~MeV$<T<$1000~MeV are presented in Table 5 of
Ref.~\cite{q-lattice}. At the highest available temperatures one
observes the ideal gas behavior, $p\cong \varepsilon/3$, but
constant $\sigma \cong \varepsilon/T^4$ is about 20\% smaller than
the SB constant $\sigma_{SB}\cong 15.63$. This behavior is similar
to the case of SU(3) gluodynamics. At small temperatures,
$T=100\div140$~MeV, the LR are expected to be smoothly connected
with the thermodynamical functions of the hadron-resonance gas.
%
\begin{figure}[ht!]
\epsfig{file=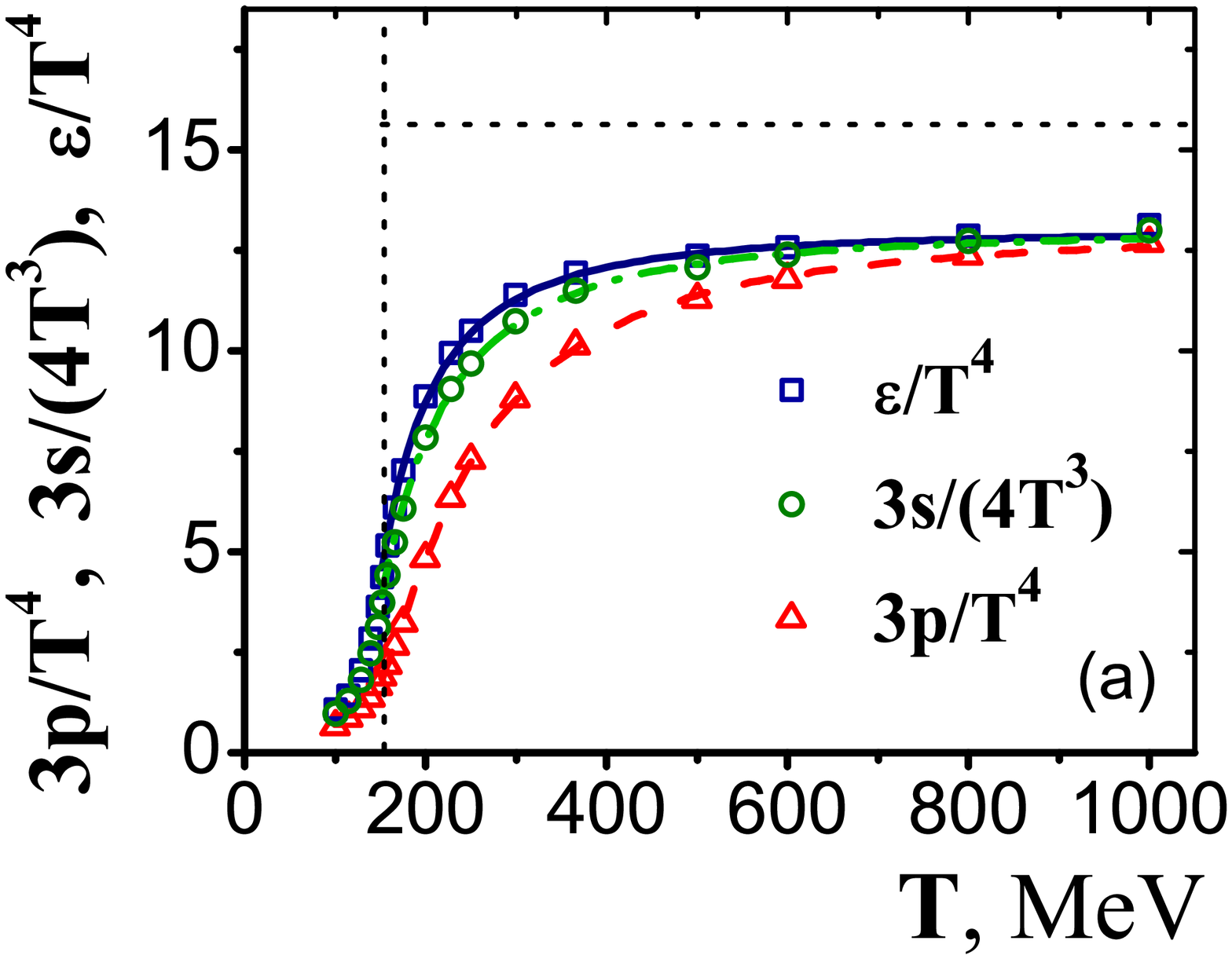,width=0.49\textwidth}
\epsfig{file=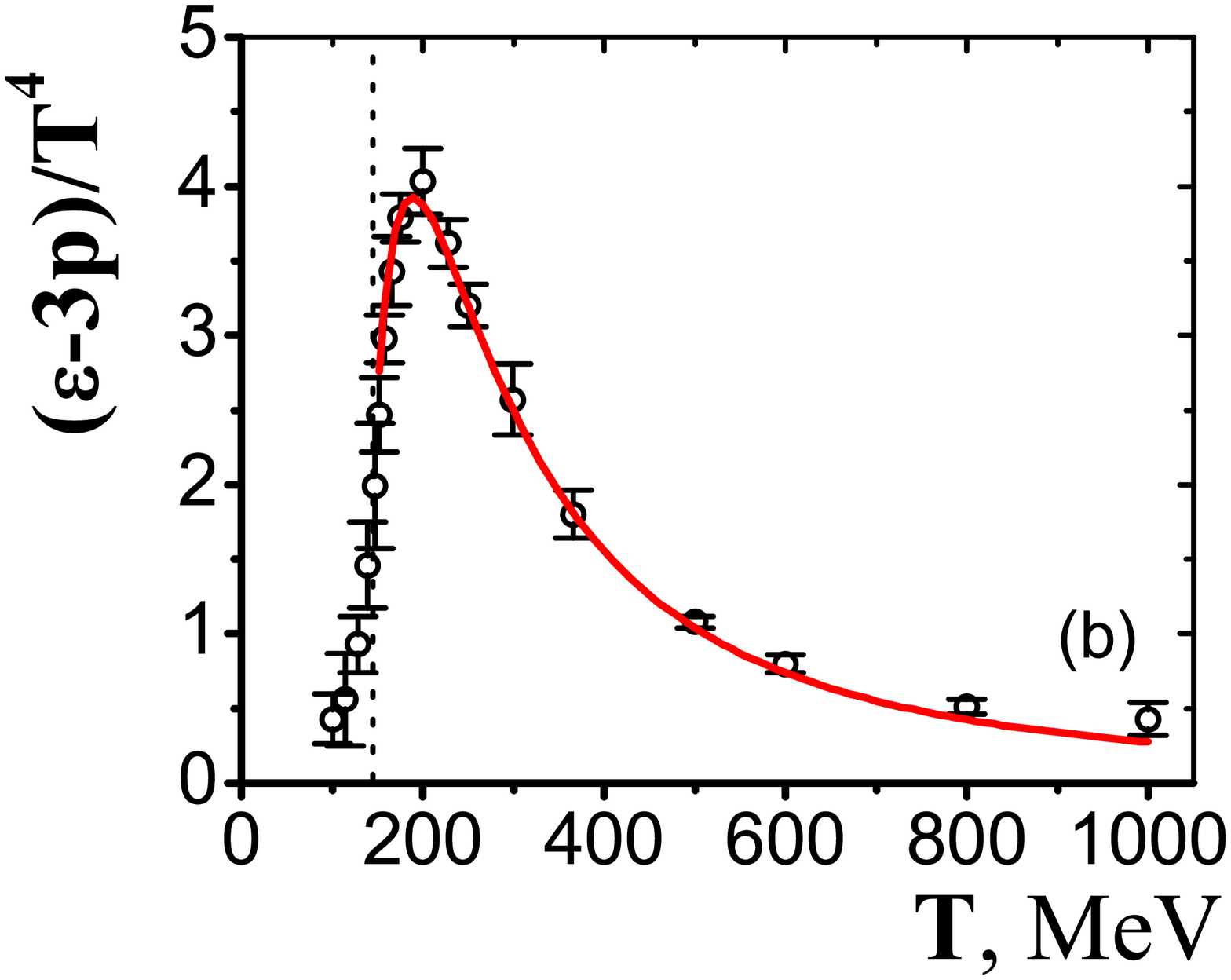,width=0.49\textwidth}\\
\epsfig{file=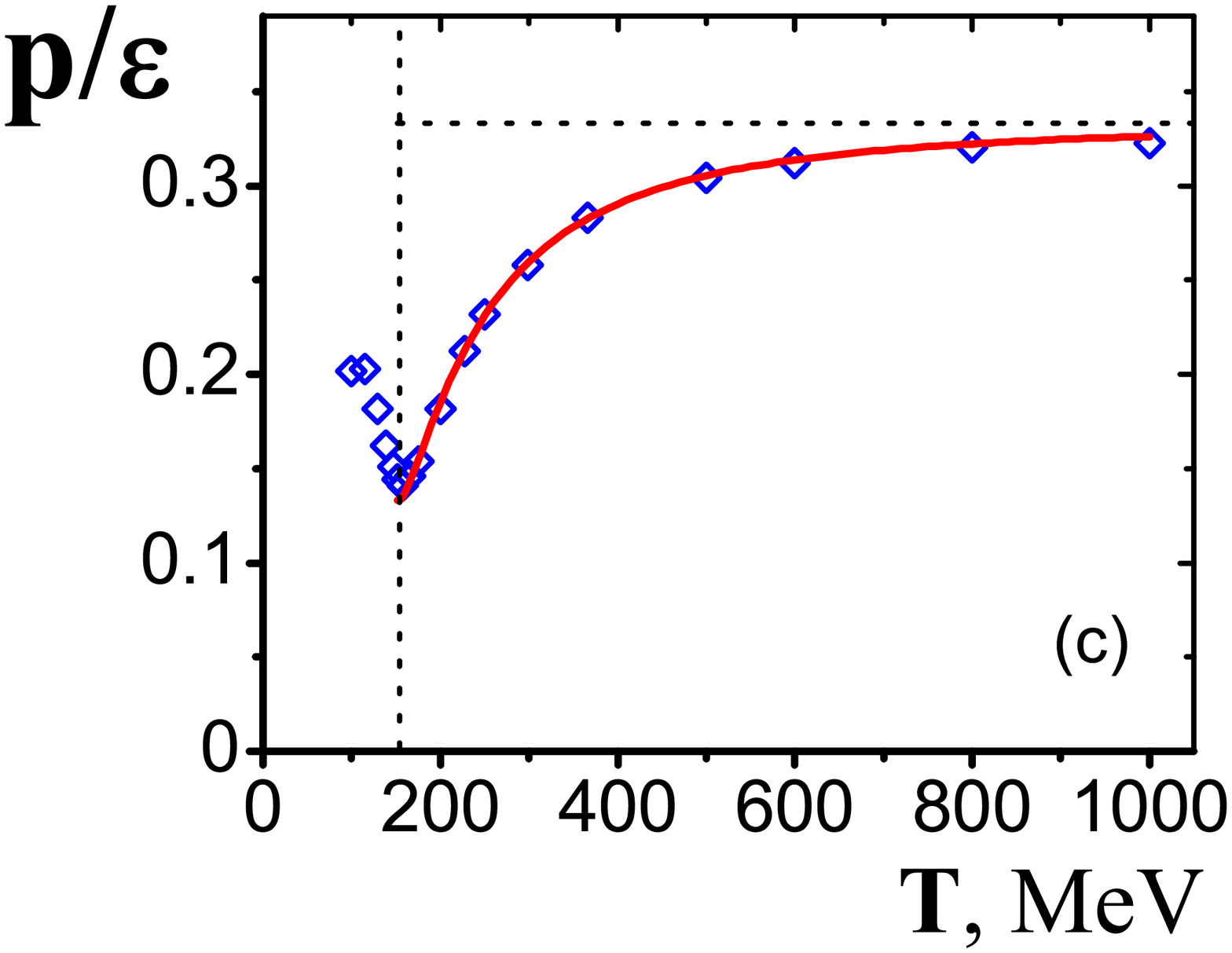,width=0.49\textwidth}
\epsfig{file=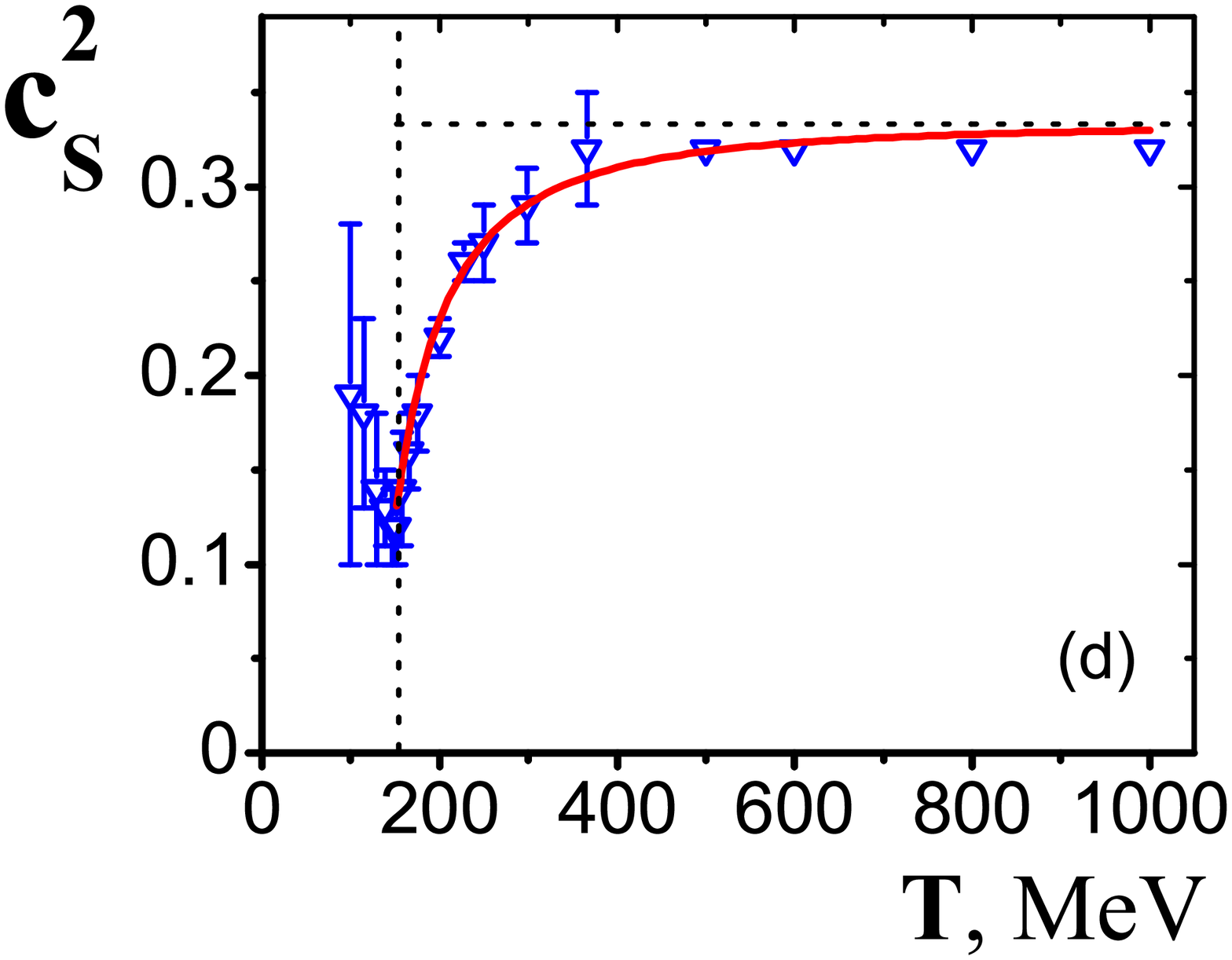,width=0.49\textwidth}
\caption{(Color online) The symbols are the MC LR for the $2+1$
QCD EoS \cite{q-lattice}.  The vertical dotted lines show
$T=T_i$=152~MeV and horizontal lines show the SB constant
$\sigma_{SB}\simeq 15.63$ in figure {\bf a}, $p/\varepsilon=1/3$
and $c_s^2=1/3$ in figures {\bf c} and {\bf d}, respectively. The
lines show the C-BM (\ref{CBM}) results at $T> T_i$=152~MeV. The
model parameters are $\sigma=13.01$, $C=6.06~T_i^2$, and
$B=-2.34~T_i^4$.~~ {\bf a}: The squares are $\varepsilon/T^4$,
triangles $3p/T^4$, and circles $3s/(4T^3)$.~~ {\bf b}:~The
interaction measure $(\varepsilon - 3p)/T^4$.~~ {\bf c}:~The ratio
$p/\varepsilon$.~~ {\bf d}:~ The speed of sound squared $c_s^2$.
\label{fig-LRQ} }
\end{figure}

In contrast to the pure SU(3) gluodynamics with a 1$^{st}$ order
phase transition between glueballs and gluons, the transition from
hadrons to quarks and gluons is a crossover.  This smooth
transition takes place in the narrow temperature range,
$T=150\div200$~MeV where the energy density increases strongly.
Several characteristic temperature points of the crossover
transition are presented in Ref.~\cite{q-lattice}:~
 $T$=145(5)~MeV at the minimum value of $c_s^2(T)$,~
 $T\equiv T_i$=152(4)~MeV at the inflection point of $(\varepsilon -
 3p)/T^4$,~
 $T$=159(5)~MeV at the minimum value of $p/\varepsilon$,~
 $T\equiv T_{max}$=191(5)~MeV at the maximum of $(\varepsilon - 3p)/T^4$.
Non of these temperatures is the critical one, and the model fit
of the QGP thermodynamical functions does not depend too much on
the choice of particular starting point in the range
$T=150\div200$. However, the hadron-resonance gas expected at low
temperatures gives a concave shape of the interaction measure
$(\varepsilon - 3p)/T^4$, while the LR \cite{q-lattice} show a
convex shape near the maximum at $T\equiv T_{max}$=191~MeV. Thus
we use the LR \cite{q-lattice} above the inflection temperature
$T>T_i$=152~MeV in our model analysis\footnote{The precise
matching of the hadron-resonance gas and LR is beyond the scope of
this paper. The discussion of a possible procedure can be found in
Ref.~\cite{HRG}.}.  This temperature will be also used to present
the model parameters.


We start with the AC-BM (\ref{ABC}) to fit the $3p/T^4$ and
$\varepsilon/T^4$ LR for the high temperature QGP phase.
The best fit corresponds to negligible values of the linear
temperature term, $A/T_i^3<< 1$. Thus, in contrast to our analysis
of the LR in the pure SU(3) gluodynamics, the AC-BM (\ref{ABC}) is
reduced to the C-BM EoS (\ref{CBM}) for QGP LR. The found model
parameters are equal to: $\sigma=13.01$, $C=6.06~T_i^2$, and
$B=-2.34~T_i^4$. A comparison of the  C-BM (\ref{CBM}) with LR
\cite{q-lattice} at $T>152$~MeV is shown in Fig.~\ref{fig-LRQ}. It
demonstrates a good agreement of the  C-BM (\ref{CBM}) with the LR
for the QGP. In particular, the model leads to the maximum
position $T_{max}\cong 189$~MeV and the value of $(\varepsilon
-3p)/T^4_{max}\cong 4$ which are very close to the LR.

%

\section{Summary}

We have considered the modifications of the bag model EoS. They
are constructed to satisfy the qualitative features expected for
the QGP EoS. We make also the quantitative comparisons with the MC
lattice results for the SU(3) gluon plasma \cite{lattice} and for
high temperature equation of state with 2+1 dynamical quarks
\cite{q-lattice}. Our modification of the bag model equation of
state includes the following features:
a suppression of the Stephan-Boltzmann constant;
linear or quadratic in temperature term in the pressure function;
a negative sign of the bag constant. These features are needed to
describe the lattice data. The best fit of the LR for
thermodynamical functions in SU(3) gluon plasma are found within
the A-bag model (\ref{ABM}). This model corresponds to:
$\varepsilon=\sigma T^4+B$, $p=\sigma T^4/3\,-AT\,-B$. A linear in
$T$ term in the pressure function is admitted by the
thermodynamical relation (\ref{therm}) between $\varepsilon(T)$
and $p(T)$. The
expression for the energy density looks formally the same as in
the standard bag model (\ref{BM}). A principal difference from the
standard bag model is a {\it negative} value of the bag constant
$B$~.

The quantitative comparison with the MC lattice results for high
temperature equation of state with 2+1 dynamical quarks
\cite{q-lattice} shows the best fit of the thermodynamical
functions for the QGP within the C-bag model (\ref{CBM}):
$\varepsilon=\sigma T^4 -CT^2+B$, $p=\sigma T^4/3\,-CT^2-B$. This
model also requires $B<0$ to fit the lattice data. Note that a
negative value of $B$ found for the gluon plasma and QGP
does not contradict to the bag model hadron
spectroscopy \cite{BM} which requires $B>0$  at zero temperature. 

\vspace{0.5cm} {\bf Acknowledgments.}~~ We thank M.~Ga\'zdzicki,
W.~Greiner, V.P.~Gusynin, P.~Huovinen,  L.L.~Jenkovszky, O.~Linnyk,
O.~Kaczmarek, E.~Megias, L.M.~Satarov, H.~Satz, and
Y.~Schr\"{o}der for fruitful discussions. V.V.~Begun thanks the
Alexander von Humboldt Foundation for support. This work was in
part supported by the Program of Fundamental Research of the
Department of Physics and Astronomy of NAS, Ukraine.



%

\end{document}